\documentstyle[12pt]{article}
\begin{document}

\title{ \Large { { \bf Path Integral Representation for Weyl Particle Propagator } } }
\author{I. A. Junior}
\maketitle

\begin{center}
Departamento de F\'\i sica, Universidade Estadual de Londrina\\
P.O. Box 6001, 86051-970, Londrina, PR, Brazil.
\end{center}

\begin{abstract}
We present a path integral representation for massless spin one-half particles. It is shown that this gives us a super--symmetric, $P-$ and $T-$non--invariant 
pseudoclassical model for relativistic massless spinning particles. Dirac quantization of this model is considered.
\end{abstract}

\vspace{0.5cm}

PACS \, $ 03.70 $ ; $ 11.10.Kk $; $ 11.15.Kc $ ; $ 03.65.Ca $

\vspace{0.5cm}

\newpage

{\bf I.\,Introduction}

\vspace{0.5cm}

In this letter we derive a path integral representation for the propagator of massless spin$-\frac{1}{2}$ particles(Weyl particles). They can be described in terms 
of a two-component spinor field $\Psi(x)$ satisfying Weyl equation,

\begin{equation}
 \label{peter}
  \imath \frac{\partial}{\partial t} \Psi(x) = - \imath \, {\bf \sigma \cdot \nabla } \Psi(x) 
   \end{equation}  
where ${\bf \sigma }$ are Pauli matrices satisfying $ \{ \sigma^{i} , \sigma^{j} \} = 2 \delta^{ij} \, (i=1,2,3)$. Path integral representation for the propagator 
of massive spinning particles  in $3+1$-dimensions has been derived in [1] where Dirac propagator has been presented as a path integral of an exponent of the pseudoclassical action for the relativistic spinning particle [2]. Path integral for the odd-dimensional case has been proposed in [4]. It has been on the even 
(in $\gamma$-matrices) form of Dirac operator. It was shown recently [11] that a similar path integral can be constructed for the even-dimensional case also.
In reference [3] the transition amplitude for massless spinning particles with $N$ extended supersymmetries has been calculated using the BRST-BFV technique. 
For $N=1$ feynman propagator for massless particles $ p^{\mu} \gamma_{\mu} / p^{2} $ has been obtained. It is a Green function for Dirac operator, but does not
satisfy Weyl condition. The path integral quantization of a pseudoclassical model for massless spinning particles considered in [5] yields the Feynman propagator
$ p^{\mu} \gamma_{\mu} ( 1 - \gamma^{5} ) / p^{2} $. This is a covariant expression; however, one encounters ordering problems. The trick used in [1], in order to 
make Dirac operator homogeneous in the $\gamma$-matrices does not work here, and the formal operator quantization of the model( following Dirac [6] ) leads to the
 equation $ \imath \, \partial_{\mu} \gamma^{\mu} (\gamma^{5} - \alpha) \Psi(x) = 0 $ for state vectors, which is not equivalent to the set consisting of Dirac
equation $ \imath \, \partial_{\mu} \gamma^{\mu} \Psi(x) = 0 $ and Weyl condition $ ( \gamma^{5} - \alpha ) \Psi(x) =0 $ , $ \alpha = \pm 1 $ . On the other hand,
we can derive Dirac equation and the Weyl condition from the operator ( Dirac and canonical [9] ) quantization of the pseudoclassical model proposed in [7]. 
So, the problem of the path integral representation for Weyl propagator is important at least for two reasons: such a representation allows us to derive a 
pseudoclassical action of a relativistic spinning particles associated with the propagator of the correspondent quantum field theory and, second, it fills the gap
in the path integral quantization of pseudoclassical mechanics.

\vspace{0.5cm}

{\bf II.\, Path integral representation}
   
\vspace{0.5cm}

According to eq(1), the weyl propagator $ S^{c}(x,y) $ satisfy the equation,

\begin{equation}
 \big( \imath \, \frac{\partial}{\partial t} + \imath \, {\bf \sigma \cdot \nabla } \big) S^{c}(x,y) = - \delta^{4}(x-y)
\end{equation}
and Feynman assymptotic conditions . As is well known, the theory is not $P-$invariant. As will be shown this is related with projector for the helicity, that 
can be presented as a pseudo quantity. Following Schwinger [8], the causal Green function $ S^{c}(x,y) $ can be presented as a matrix element
$S^{c}(x,y) = <x| \hat{S}^{c} |y> $ of an operator $ \hat{S}^{c} $[8] in some Hilbert space, where eigenvectors$ |x> $of the self-conjugated coordinate 
operators $X^{\mu}$ form a complete set,

\begin{eqnarray*}
 \label{pan}
  X^{\mu} |x> = x^{\mu} |x>,\\
  \\
   <x|y> = \delta^{4}(x-y), \\
    \\
     \int \, dx \,  |x><x| = I.
\end{eqnarray*}

The operator $ \hat{S}^{c} $ satisfies the equation, 

\begin{equation}
\label{stark}
\big( \hat{P}_{0} - \sigma^{i} \hat{P}^{i} \big) \hat{S}^{c} = I,
\end{equation}
where $ P_{\mu} $ is the operator canonically conjugated to $ X^{\mu} $,

\begin{equation}
\label{romanov}
<x|P_{\mu}|y> = - \imath \, \partial_{\mu} \delta^{4}(x-y), \, P_{\mu}|p> = p_{\mu}|p>, \, \int \, dp \, |p><p| = I,
\end{equation}

\begin{equation}
\label{maria}
\big[ \hat{P}_{\mu} , \hat{X}^{\nu} \big]_{-} = - \imath \, \delta^{\nu}_{\mu} , \, <p|p'> = \delta^{4}(p - p'),
\end{equation}
and $ \hat{P}^{\mu} = \eta^{\mu\nu} \hat{P}_{\nu} $, $ \eta^{\mu\nu} = diag( 1, -1, -1, -1 ) $. The equation (2) is then considered to be a matrix element
of the operator equation $ \big( \hat{P}^{0} - {\bf \sigma \cdot \hat{P} } \big) \hat{S}^{c} = -I $, then 

\begin{eqnarray*}
\label{jardim}
\hat{S}^{c} = - \frac{ P^{0} + { \bf \sigma \cdot P } }{ P_{\mu}P^{\mu} } \equiv - \frac{ \hat{F}_{+} }{ \hat{F}_{-} \hat{F}_{+} } 
\end{eqnarray*}
where $\hat{F}_{\pm} = P^{0} \pm {\bf \sigma \cdot P} $ \, .

The $ { \bf \sigma } $ matrix can be express as $ \big( \sigma^{k}  = - \frac{ \imath }{ 2 } \epsilon^{ kij } \sigma^{i} \sigma^{j} \big) $ so, this 
can be considered as Bose-type operator. Using the Schwinger proper-time representation for an inverse operator $ \frac{1}{ \hat{F}_{-} \hat{F}_{+} } $ [1] and 
an additional representation of the operator $ \hat{F}_{+} $ , by means of a Gaussian integral over two Grassmann variables $\kappa_{1}$ e $ \kappa_{2} $ with the 
involution property $ \big( \kappa_{1} \big)^{+} = \kappa_{2} $ [4], we can write 

\begin{equation}
\label{lucimara}
\hat{S}^{c} = - \frac{ P^{0} + { \bf \sigma \cdot P } }{ P_{\mu}P^{\mu} } =
\imath \,
 \int_{0}^{\infty} \, d\lambda \, \int \, d\kappa \, e^{\imath \, \big[ \lambda \big( \hat{F}_{-}\hat{F}_{+} + \imath\epsilon \big) + \kappa \hat{F}_{+} \big] }
\end{equation}
where $ \kappa = \kappa_{1}\kappa_{2} $ , \, $d\kappa = d\kappa_{1}d\kappa_{2}$.

Thus, we get for the propagator 

\begin{equation}
\label{cinderela}
\hat{S}^{c}(x_{out},x_{in} ) = \imath
\int_{0}^{\infty} d\lambda \, \int d\kappa < x_{out}| e^{-\imath \hat{{\cal H}}(\lambda,\kappa)} |x_{in}>
\end{equation}
where $  \hat{{\cal H}}(\lambda,\kappa) = \lambda P^{2} - \kappa \big(P^{0} + {\bf \sigma \cdot P} \big)  $ . Similarly to [1], the matrix element in (7) can be
 presented by means of  a path integral ( taking into account the Weyl-ordering procedure for operators) was

\begin{eqnarray*}
\label{jose}
\hat{S}^{c}(x_{out},x_{in} )  & = &   
\imath exp \big( \imath \gamma^{\mu} \frac{\partial_{l}}{\partial \Theta^{\mu} } \big) 
\int_{0}^{\infty} \, d\lambda_{0} \, \int \, d\kappa_{0} \, \int_{\lambda_{0}} \, D\lambda \, \int_{\kappa_{0}} \, D\kappa \times \\
& \, & \\
&  \times & 
\int_{x_{in}}^{x_{out}} Dx \, \int Dp \, \int D\pi \, \int D\varrho \, \int_{\psi(0) + \psi(1) = \theta} D\psi \times \\
& \, & \\
& \times & 
exp \left\{  \imath \int_{0}^{1} 
\left[ \lambda P^{2} + \kappa \big( -2\imath \epsilon^{jkl} P^{j} \Psi^{k} \Psi^{l} + P^{0} \big) \right. \right. + \hspace{1.5cm}(8) \\
& \, & \\
& - & \left. 
\imath \psi_{\mu} \dot{\psi}^{\mu}  + p_{\mu}\dot{x}^{\mu} + \pi \dot{\lambda} + \varrho\dot{\kappa}   \right] d\tau + \psi_{\mu}(1) 
\psi^{\mu}(0) \left.  \right\} |_{\theta =0}
\end{eqnarray*}
where $ P^{\mu} = - p^{\mu} $ , $ x(\tau) $ , $ \lambda(\tau) $ and $ \pi(\tau) $ are even and $ \varrho_{1}(\tau) $ , $ \varrho_{2}(\tau) $,
$ \kappa_{1}(\tau) $ , $ \kappa_{2}(\tau) $ and $ \psi(\tau) $ are odd trajectories, obeying the boundary conditions $ x(0) = x_{in}$,
$ x(1) = x_{out} $ , $ \lambda(0) = \lambda_{0} $ , $ \kappa(0) = \kappa_{0} $ , $ \psi(0) + \psi(1) = \theta$ ($\theta^{\mu} $ are odd variables 
anticommuting with $\sigma$-matrices) and the following notations are used

\begin{eqnarray*}
\label{banana}
\kappa = \kappa_{1}\kappa_{2} \, , \, 
\varrho\dot{\kappa} = \varrho_{1}\dot{\kappa}_{1} + \varrho_{2}\dot{\kappa}_{2} 
\end{eqnarray*}
  
\begin{eqnarray*}
\label{manga}
d\kappa = d\kappa_{1}d\kappa_{2} \, , \, 
D\kappa = D\kappa_{1} D\kappa_{2} \, , \,
D\varrho = D\varrho_{1} D\varrho_{2}
\end{eqnarray*}

Introducing the four vector $ k^{\mu} = ( k,0,0,0)$ and the $ \sigma^{0}$-matrix where  $ \sigma^{0}= diag(1,1) $; integrating over momenta and making the 
substitution $ e = 2\lambda $ , we get the path integral in the Lagrangian form

\begin{eqnarray*}
\label{jamelao}
\hat{S}^{c} & = &   
\imath exp \big( \imath \sigma^{\mu} \frac{\partial_{l}}{\partial \theta^{\mu} } \big) 
\int_{0}^{\infty} \, de_{0} \, \int \, d\chi_{0} \, \int \, d\kappa_{0} \, \int_{e_{0}} \, M(e)De \times \\
& \, & \\
&  \times & 
\int_{\kappa_{0}} D\kappa \, \int_{x_{in}}^{x_{out}}  Dx \, \int Dp \, \int D\pi \, \int D\varrho \,  \int_{\psi(0) + \psi(1) = \theta} D\psi \times \\
& \, & \\
& \times & 
exp \left\{  \imath \int_{0}^{1} \right. 
\left[ - \frac{ 1}{2e}  \big( \dot{x}^{\mu} + \imath \epsilon^{\mu\nu\rho\iota} \kappa_{\nu}  \psi_{\rho} \psi_{\iota}  + \frac{1}{2} \kappa^{\mu}  \big)^{2}
\right. + \hspace{3.5cm}(9) \\
& \, & \\
& - & \left. 
\imath \psi_{\mu} \dot{\psi}^{\mu}  + \pi \frac{\dot{e}}{2} + \varrho_{\mu} \frac{\dot{\kappa}^{\mu}}{2}    \right] d\tau + \psi_{\mu}(1)\psi^{\mu}(0) 
 \left.  \right\} |_{\theta =0}
\end{eqnarray*}
where

\begin{eqnarray*}
\label{napoleao}
M(e) = \int \, Dp \, exp \left\{ \frac{\imath}{2} \int_{0}^{1} \, ep^{2} d\tau \right\}
\end{eqnarray*}
and $ S^{e}_{GF} = \int_{0}^{1} \, \pi \frac{ \dot{e} }{2} \, d\tau $, 
$ S^{\kappa}_{GF} = \int_{0}^{1} \, \varrho_{\mu}  \frac{ \dot{\kappa}^{\mu}  }{2} \, d\tau $ are gauge fixing terms.

\vspace{0.5cm}

{\bf III. \, Dirac Quantization}

\vspace{0.5cm} 

Consider now the Diarc quantization[6,9] of the action which appears in the exponent of the eqaution (9). Without the gauge fixing terms we have

\begin{eqnarray*}
\label{natalia}
S =   \int_{0}^{1}  
\left[ - \frac{ 1}{2e}  \big( \dot{x}^{\mu} + \imath \epsilon^{\mu\nu\rho\iota} \kappa_{\nu}  \psi_{\rho} \psi_{\iota}  + \frac{\alpha}{2} \kappa^{\mu}  \big)^{2}
-\imath \psi_{\mu} \dot{\psi}^{\mu}
\right] d\tau  \hspace{1cm} (10)
\end{eqnarray*} 

There are three types of gauge transformations under which this action is invariant: reparametrization $ \delta x^{\mu} = \dot{x}^{\mu} \xi $ ,
$ \delta e  = \frac{d}{d\tau} (e\xi)  $,  $ \delta \kappa^{\mu} = \frac{d}{d\tau} ( \kappa^{\mu}\xi)  $,
$ \delta \psi^{\mu} = \dot{\psi}^{\mu} \xi $ , with an even parameter $\xi(\tau)$ ; supertransformation,
$ \delta x^{\mu} = \imath \psi^{\mu} \epsilon $,
$ \delta \kappa^{\mu} = 0 $,
$ \delta \psi^{\mu} = \frac{1}{2e} z^{\mu} \epsilon $,
$ z^{\mu} = \dot{x}^{\mu} + \imath \epsilon^{\mu\nu\rho\varsigma} \kappa_{\nu}\psi_{\rho} \psi_{\varsigma} + \frac{1}{2} \kappa^{\mu} $,
with an odd parameter $ \epsilon(\tau) $; and the gauge transformation
$ \delta e = 0 $,
$ \delta \kappa^{\mu} = \frac{d}{d \tau} (\kappa^{\mu} \phi)  $,
$ \delta \psi^{\mu} = \frac{1}{e} \epsilon^{\mu\nu\rho\varsigma} \kappa_{\nu}z_{\rho} \psi_{\varsigma} \phi   $,
$ \delta x^{\mu} = - \big( \epsilon^{\mu\nu\rho\varsigma } \kappa_{\nu} \psi_{\rho} \psi_{\varsigma} + \frac{\alpha}{2} \kappa^{\mu} \big) \phi  $
, with an even parameter $ \phi(\tau) $ . The equations of motions have the form,

\begin{eqnarray*}
\label{dilane}
\frac{\delta S}{\delta x^{\mu} } &=& \frac{d}{d\tau} \left[ \frac{1}{e} z_{\mu} \right] = 0 \\
\\
\frac{\delta S}{\delta e } &=&  \frac{1}{2e^{2}} z^{2}  = 0 \\
\\
\frac{\delta S}{\delta \kappa^{\mu} } &=& -\frac{1}{e}  z^{\nu}  \left[\imath 
\epsilon_{\nu\mu\rho\varsigma }  \psi^{\rho} \psi^{\varsigma}  -\frac{\imath \alpha}{2} \eta_{\mu\nu} \right] = 0 \hspace{5cm} (11)\\
\\
\frac{\delta_{r} S}{\delta \chi  } &=&  \frac{\imath}{e} z_{\mu}\psi^{\mu}  = 0 \\
\\
\frac{\delta_{r} S}{\delta \psi^{\mu}  } &=&  2\dot{\psi}_{\mu} -  \frac{1}{e} z^{\rho} 
\big( \eta_{\mu\rho}\chi - 2 \epsilon_{\mu\rho\nu\varsigma} \kappa^{\nu} \psi^{\varsigma} \big)   = 0 
\end{eqnarray*}

Going over to the Hamiltonian formulation, we introduce the canonical momenta: 
$ p_{\mu} = \frac{\partial L}{\partial\dot{x}^{\mu}} = - \frac{1}{e} z_{\mu} $,
$ P_{e} = \frac{\partial L}{\partial\dot{e} } = 0 $,
$ P_{\psi^{\mu}}  = \frac{\partial_{r}  L}{\partial\dot{\psi}^{\mu}} = - \imath \psi_{\mu}  $,
$ P_{\kappa_{\mu}}  = \frac{\partial L}{\partial\dot{\kappa}^{\mu}} = 0 $, and follow of this the primary constraints
$ \Phi_{1}^{(1)}  = P_{e}  $,
$ \Phi_{2\mu}^{(1)}  = P_{\mu} + \imath \psi_{\mu} $,
$ \Phi_{3\mu}^{(1)} = P_{\kappa_{\mu}}  $. The total Hamiltonian $ H^{(1)} = H + \lambda_{A} \Phi_{A}^{(1)} $ , can be constructed according to the standard 
procedure which gives

\begin{eqnarray*} 
\label{touro}
H = -\frac{e}{2} p^{2} + \imath 
\big( \epsilon_{\nu\mu\rho\varsigma} p^{\mu}\psi^{\rho} \psi^{\varsigma} + \frac{\imath\alpha}{2} p_{\nu} \big) \kappa^{\nu} \,\,\,\,\,\,\, (12)
\end{eqnarray*} 

From the conditions of the conservation of the primary constraints in time $ \tau $, $ \dot{\Phi}^{ (1) } = \left\{ \Phi^{ (1) } , H^{ (1) } \right\} = 0$ , we
find the set of independent secondery constraints
$ \Phi_{1}^{(2)}  = p^{2} $ , 
$ \Phi_{2\mu}^{(2)}  = T_{\mu} = 
\epsilon_{\mu\nu\rho\varsigma} p^{\nu}\psi^{\rho} \psi^{\varsigma} + \imath \frac{1}{2} p_{\mu}$, and determine $\lambda$ , which correspond to the primary 
constraint $ \Phi_{1}^{(2)} $ .  One can go over from the initial set of independent constraints $( \Phi^{(1)}, \Phi^{(2)} )$ to the equivalent one 
 $( \Phi^{(1)}, \tilde{\Phi}^{(2)} )$, where $ \tilde{\Phi}^{(2)} = \Phi^{(2)} |_{\psi \to \tilde{\psi} = \psi + \frac{\imath}{2} \Phi_{2}^{(1)}} $.
The new set of constraints can be explicitly divided in a set of first-class constraints, which are $ ( \Phi_{1}^{(1)} , \Phi_{3}^{(1)} , \tilde{\Phi}^{(2)})$ 
and in a set of second-class constraints , $ \Phi_{2}^{(1)} $ .

Let consider the Dirac quantization of the action(10) where the second-class constraints $\Phi_{2}^{(1)}$ define the Dirac brackets and commutation relations.
It is necessary to remember that the quadrivector $ \kappa^{\mu} = (1,0,0,0) $ was defined in one particular referencial frame so, this meant that the set of 
second-class constraint $\Phi_{2}^{(1)} $ has in fact only the three constraint $\Phi_{2i}^{(1)} \Phi_{2j}^{(1)} \Phi_{2k}^{(1)} $. The first-class 
constraints, being  applied to the state vectors, define the physical states.
For essential operators and non zeroth commutators we have 
$ \left[ \hat{x}^{\mu} , \hat{p}_{\nu} \right]_{-} = \imath \left\{ x^{\mu} ,p_{\nu} \right\}_{D(\Phi_{2}^{(1)})} = \imath \delta^{\mu}_{\nu} $,
 $ \left[ \hat{\psi}^{i} , \hat{\psi}_{j} \right]_{+} = \imath \left\{ \psi^{i} ,\psi^{j} \right\}_{D(\Phi_{2}^{(1)})} = \frac{1}{2}  \delta^{i}_{j} $.
It is possible to construct a realization of these commutation relation in a Hilbert space ${\cal R}$ whose elements $\Psi$ are 2-components columns. Then 
$ \hat{\psi}^{i} = \frac{1}{2} \sigma^{i}$, 
$ \hat{x}^{\mu} = x^{\mu} I $,
$ \hat{p}_{\mu} = - \imath \partial_{\mu} $ . 
Using the primary first-class constraints operators $ \hat{\Phi}^{(1)}_{1,3} $ in this way, one can see that physical vectors are only functions on $x$. 
The operators of the secondary first-class constraints $\hat{\Phi}^{(2)} $, being applied to the state vectors, give the equations
$ \hat{p}^{2}\Psi(x) = 0 $, 
$ \hat{p}_{\mu}\gamma^{\mu} \Psi(x) = 0 $,
$ \hat{T}_{0}\Psi(x) = 0 $,
$ \hat{p}^{0}\Psi(x) = {\bf \sigma \cdot \hat{p}} \Psi(x) $,
which are just the Dirac equation for massless particles and Weyl condition. The canonical quantization(with gauge fixing [9] can also be done and give 
the same quantum mechanics.

\vspace{0.5cm}

{\bf IV. Conclusions}

\vspace{0.5cm}

In this paper, we have presented the path integral formulation for weyl particle propagator with the help of the helicity of the bosonic-type operator.
The pseudoclassical action (10) also has been derivated. it is shown that a general formula for an operator 
$ \hat{S}^{c} = -\frac{\hat{F}}{\hat{F}^{2}} \hat{C}  $ can be written in the form (6), where $\hat{F} $ is the fermionic operator, 
$ \hat{F}^{2} $ is the bosonic one and , $ \hat{C} $ is a bosonic-type operator( of an even degree in the $\gamma$-matrices).

The pseudoclassical action (10) is not  a Lorentz covariant since $\kappa^{\mu} $ is a time-like four-vector; however 
their quantization leads to the Lorentz covariant quantum mechanical model which is equivalent to the theory of weyl particles.
On the other hand, the quantization of the Lorentz covariant pseudoclassical action (10) (with the four-vector 
$\kappa^{\mu} \neq (\kappa,0,0,0) $ leads to the same Lorentz covariant quantum mechanics of Weyl particles, but , in this case, not all constraints 
$ \hat{T}_{\mu} = p_{\mu} T /p^{0} $ are independent[10]

\vspace{0.5cm}

{\bf  \, Acknowledgements}
 
\vspace{0.5cm}

The author are indebted to A. E. Gon\c{c}alves for several valuable observations about this work and CAPES by financial support.


\vspace{0.5cm}
 

 

\begin{thebibliography}{99}
\bibitem{fragit}E.S. Fradkin and D.M. Gitman, Phys. Rev. D
{\bf 44}, 3220 (1991).
\bibitem{ber2}F.A. Berezin and M.S. Marinov, Pis'ma Zh. Eksp. Theor.
Fiz. {\bf 21}, 678 (1975) [JETEP] Lett. {\bf 21}, 320 (1975)];
Ann. Phys. (N.Y.) {\bf 104}, 336 (1977).
\bibitem{riva}M. Pierri and V.O. Rivelles, Phys. Lett. B {\bf 251},
421 (1990).
\bibitem{git} D.M. Gitman, Nucl. Phys. B {\bf 488}, 490 (1997) and
references therein.
\bibitem{bcd}A. Barducci, R. Casalbuoni, D. Dominici and
L. Lusanna, Phys. Lett. B {\bf 100}, 126 (1981).
\bibitem{dir}P.A.M Dirac, {\em Lectures on Quantum Mechanics}
(Yeschiva University, 1964).
\bibitem{ggt}D.M. Gitman, A.E. Gon\c calves and I.V. Tyutin, Phys. Rev. D
{\bf 50}, 5439 (1994); Preprint IFUSP/P-1094, dezember/1993.
\bibitem{sch}J. J. Schwinger, Phys. Rev. {\bf 82}, 664 (1959).
\bibitem{gt}D.M. Gitman and I.V. Tyutin {\em Quantization of Fields
withs Constraints} (Springer, Berlin, 1990).
\bibitem{gg}D.M. Gitman and A.E. Gon\c calves, Int. J. Theor. Phys.
{\bf 35}, 2459 (1996).
\bibitem{zla}S.I. Zlatev, Phys. Rev. D, (2000)
\end{thebibliography}
\end{document}